\documentclass[runningheads]{llncs}

\usepackage{graphicx}
\usepackage{amsmath}

\begin{document}

\title{End-to-end lung nodule detection framework with model-based feature projection block.
\thanks{Supported by BOTKIN.AI, Skolkovo}
% \thanks{Supported by Anonymous Organization}
}

\titlerunning{End-to-end lung nodule detection framework}

\author{Ivan Drokin\inst{1,2}\orcidID{0000-0003-2634-7857} \and
Elena Ericheva\inst{1,3}\orcidID{0000-0002-9460-634X}}
\authorrunning{I. Drokin et al.}
\institute{Intellogic Limited Liability Company (Intellogic LLC), office 1/334/63, building 1, 42 Bolshoi blvd., territory of Skolkovo Innovation Center, 121205, Moscow, Russia \and
\email{ivan.drokin@botkin.ai} \and
\email{elena.ericheva@botkin.ai}}

% \author{Anonymous\inst{1,2}\orcidID{****-****-****-****}}
% \authorrunning{Anonymous}
% \institute{Anonymous Organization \and
% \email{**@******.***}}

% \author{Ivan Drokin\inst{1,2}\orcidID{0000-0003-2634-7857}}
% \authorrunning{I. Drokin et al.}
% \institute{Intellogic Limited Liability Company (Intellogic LLC), office 1/334/63, building 1, 42 Bolshoi blvd., territory of Skolkovo Innovation Center, 121205, Moscow, Russia \and
% \email{ivan.drokin@botkin.ai}}

\maketitle              

\begin{abstract}
This paper proposes novel end-to-end framework for detecting suspicious pulmonary nodules in chest CT scans. The method’s core idea is  a  new  nodule  segmentation  architecture  with  a model-based feature projection block on three-dimensional convolutions.  This  block  acts  as  a  preliminary  feature  extractor  for  a two-dimensional  U-Net-like  convolutional  network.  Using  the  proposed approach along with an axial, coronal, and sagittal projection analysis makes it possible to abandon the widely used false positives reduction step. The proposed method achieves SOTA on LUNA2016 with 0.959 average sensitivity, and 0.936 sensitivity if the false-positive level per scan is 1/4. The paper describes the proposed approach and represents the experimental results on LUNA2016 as well as ablation studies. 
% The code of the proposed model is available at https://github.com/******.

\keywords{Image analysis \and Computer vision \and Segmentation \and Computer detection \and Computer-Assisted \and Lung cancer \and Pulmonary nodule \and False-Positive Reduction \and Chest CT. }
\end{abstract}

\section{Introduction and previous work}

Survival in lung cancer (over five years) is approximately 18.1\%\footnote{2018 state of lung cancer report: https://www.naaccr.org/2018-state-lung-cancer-report/}. 
Early-stage lung cancer (stage I) has a five-year survival rate of 60-75\%. A recent National Lung Screening Trial (NLST) study revealed that lung cancer mortality can be reduced by at least 20\% by using a high-risk annual screening program with low-dose computed tomography (CT) of the chest \cite{Introduction_2}. Computerized tools, especially image analysis and machine learning, are critical factors for improving diagnostics, facilitating the identification of results that require treatment, and supporting medical experts’ workflow \cite{Introduction_6}. Computer-aided detection (CAD) systems have also shown improvements in radiologists’ readability \cite{Introduction_8,Introduction_9,Introduction_10}. 

Modern CAD systems are built using a deep learning approach to solve the detection task. CNN models are widely studied. Due to their specificity, CNNs can work efficiently with images and focus on candidate recognition \cite{Introduction_6}. 
A multicontextual 3D residual convolutional neural network (3D Res‐CNN) is proposed in \cite{Prev_15a,Prev_15b}. 
The method presented in \cite{Prev_15c} is based on structural relationship analysis between nodule candidates and vessels.
In \cite{Prev_15d}, a rule-based classifier is used to eliminate apparent non-nodules, followed by a multi-view CNN. 
The CNN from \cite{Prev_15e} is fed with nodule candidates obtained by combining three candidate detectors specifically designed for solid, subsolid, and large nodules. For each candidate, a set of 2D patches from differently oriented planes is extracted. 

Currently, the conventional pipeline in the screening task for CAD consists of several stages — principally detection and cancer classification. A two-stage machine learning algorithm is a popular approach that can assess the risk of cancer associated with a CT scan \cite{Prev_12a,Prev_12b,Prev_13,Prev_14a,Prev_14b}. The first stage uses a nodule detector, which identifies nodules contained in the scan. The second step is used to assess whether nodules are malignant or benign. An evaluation of CAD systems is performed on independent datasets from the LUNA16\footnote{LUNA16 challenge homepage: https://luna16.grand-challenge.org/} and ANODE09\footnote{ANODE09 challenge homepage: https://anode09.grand-challenge.org/} challenges and the DLCST\cite{Prev_16} dataset.

In this work, we have proposed an end-to-end approach for both stages. The detector stage and false-positive reduction stage were developed and inferred in one convolutional neural network. The proposed novel architecture is based on a trainable analog of the Maximum Intensity Projection (MIP) \cite{med_mip0} block, which acts as a preliminary feature extractor, followed by a U-Net-like segmentation network. This end-to-end approach doesn't require any specific data sources or markups but does demand some data preparation. We have described the proposed data processing and augmentation techniques, and an evaluating process that includes current SOTA comparisons.

\section{Method}
\label{sec:Method}
\subsection{Overall pipeline description}
\label{subsec:Overall pipeline description}
\begin{figure}[tbp]
\centering
  \includegraphics[width=0.8\linewidth]{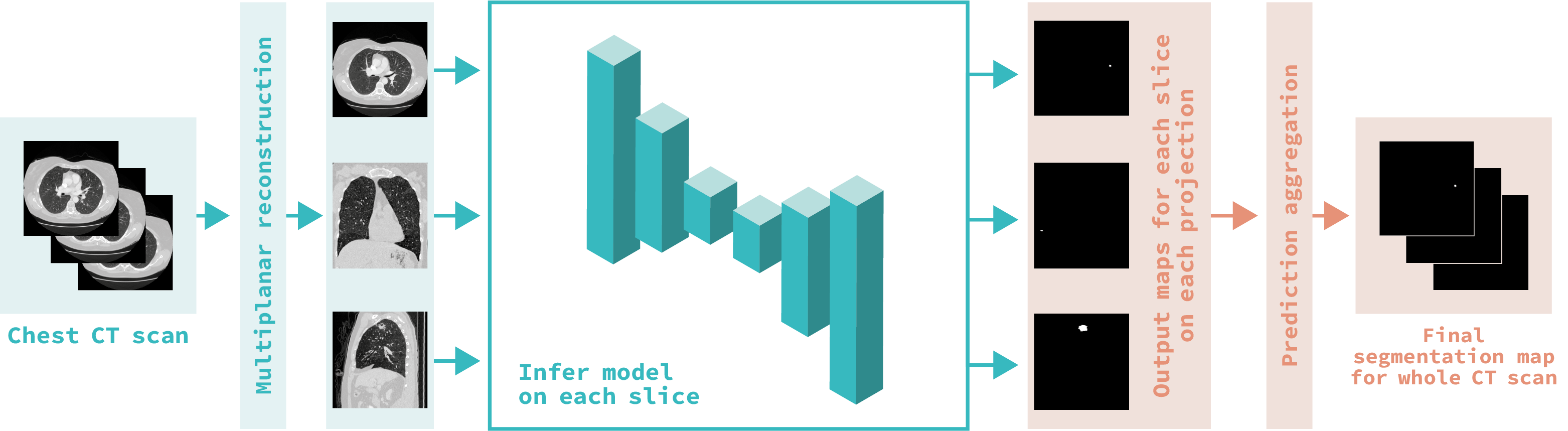}
  \caption{Scheme of the proposed framework for lung nodule detection}
  \label{fig:Overall}
\end{figure}
A high-level scheme of our solution is presented in Figure \ref{fig:Overall}. We propose a framework based on semantic segmentation for lung nodule detection. As a model input, we apply not only widely used axial images from the CT study but also sagittal and coronal projections. Radiologists often use multiplanar reconstruction to obtain more accurate results from complex findings analysis. For example, through the use of analysis types such as differentiation between plane and volume objects (e. g., fibrosis and nodules) and the examination of the relation of those findings to adjacent structures such as bronchi and vessels \cite{med_mpr}. We use data from all three projections simultaneously and obtain a single model to detect nodules, analyzing any given projection during the training procedure. This approach enables us to train a more robust model without expanding the training dataset. During the inference procedure, we first prepare sagittal and coronal projections of a CT study. After that, we infer the model on each slice independently. Then, we average prediction by applying the reverse data transformation to the axial projection. This approach offers us to obtain a more accurate representation of the three-dimensional shape of the findings using two-dimensional convolutional networks and, as a consequence, reduces false positives. Along with the redesigned segmentation network architecture, this framework's construction allows us to eliminate the false-positive reduction step.

The segmentation network itself consists of two blocks. The first block is a trainable analog of MIP \cite{med_mip0}, which acts as a preliminary feature extractor, followed by a U-Net-like segmentation network. This approach offers the advantages of both a 3D decoder and a 2D one. Conv3D-based models bring high-quality results. The U-Net-like network enables us to train and infer using a reasonable amount of computing resources.

\subsection{Model-based feature projection}
\label{subsec:Model-based Maximum Intensity Projection}
\begin{figure}[tbp]
\centering
  \includegraphics[width=0.8\linewidth]{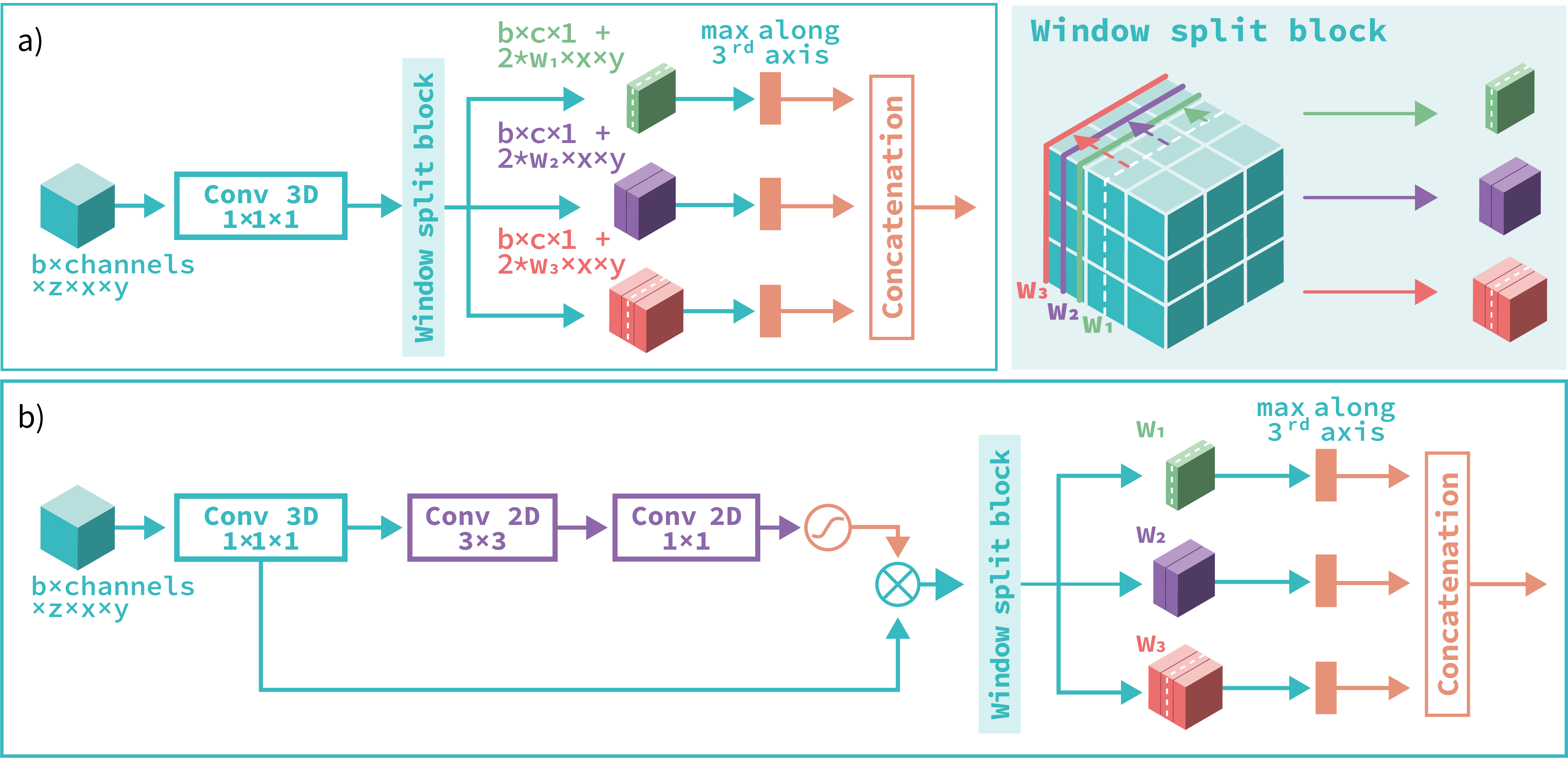}
  \caption{Scheme of the proposed model-based feature projection. a. Base block design with 3D convolutional bottleneck and maximum feature aggregation. b. Attention-like block design with additional self-attention block}
  \label{fig:mbmip}
\end{figure}

MIP is a 2D projection of voxels with the highest attenuation value on every view throughout the volume. This method tends to display bone and contrasting material-filled structures preferentially. Other lower-attenuation structures are not well visualized. The primary clinical application of MIP is to improve the detection of pulmonary nodules and assess their profusion. MIP also helps characterize the distribution of small nodules. In addition, MIP sections of variable thickness are excellent for evaluating vessels' size and location, including the pulmonary arteries and veins \cite{med_mip1}. 

Zheng et al. \cite{ml_mip} proposed to use MIPs as the input data for a detection model in a lung nodule screening pipeline. This solution led to more accurate models but still requires a false positive reduction step. We propose to extend the MIP concept to a trainable feature extractor block with 3D-convolutions as the primary step, followed by maximum feature aggregation. 

\label{subsec:Segmentaion network with model-based Maximum Intensity Projection blocks}
\begin{figure}[tbp]
\centering
  \includegraphics[width=0.8\linewidth]{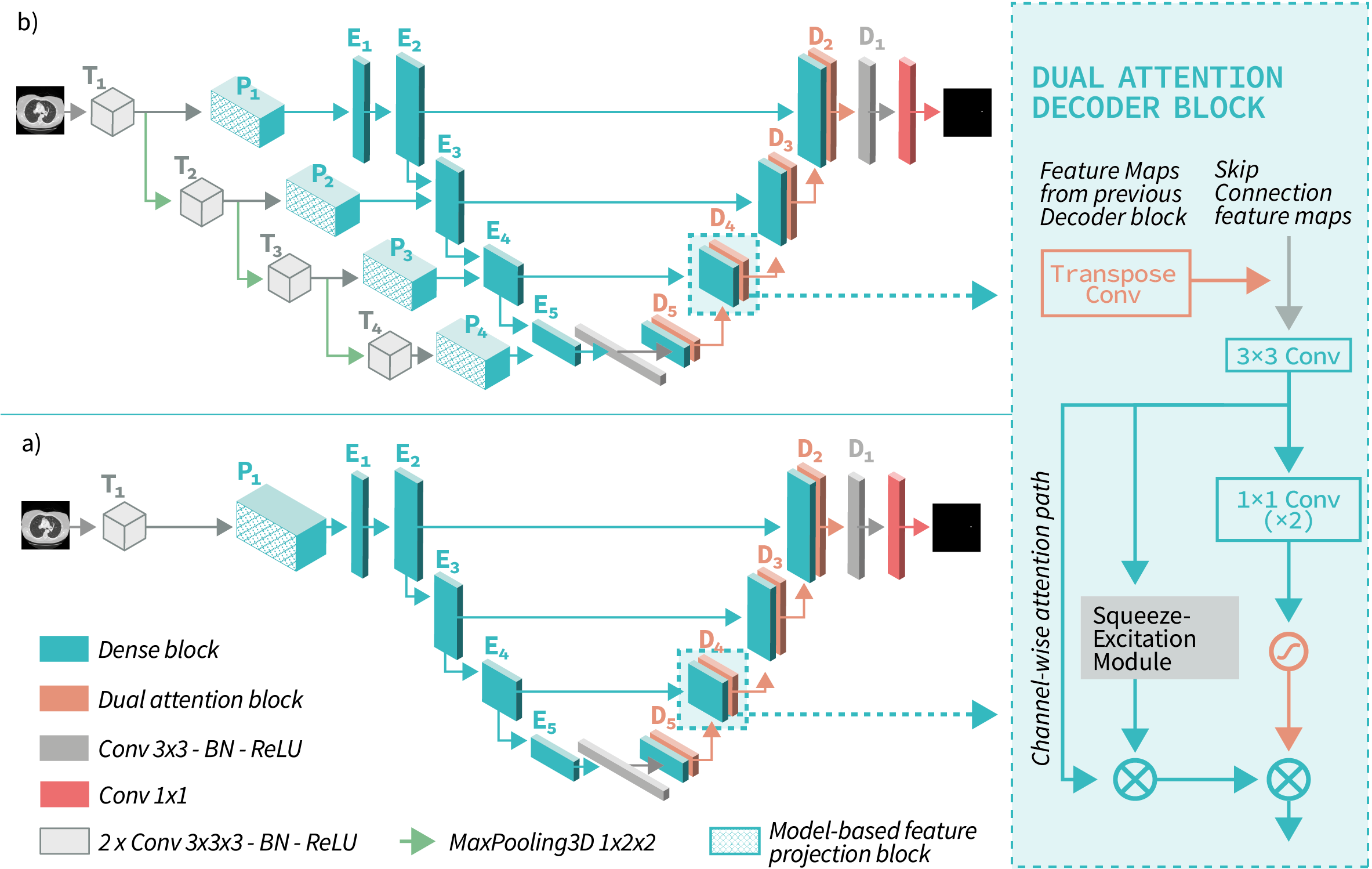}
  \caption{Scheme of the proposed segmentation model architectures.}
  \label{fig:mbnet}
\end{figure}

Let's denote a chest CT scan as $D \in R^{d \times n \times n}$. The first axis is aligned with the patient's spine. The two remaining axes are the width and height of the image. Thus, the slab for a given position $p$ and width $w$ is $S(p, w) = \{D(i, j, k) | p=p-w \dots p+w; j,k=1 \dots n\}$. We apply two consecutive blocks of a 3D convolution layer with kernel size $3 \times 3 \times 3$, batch normalization, and ReLU activation to slab $S(p, w)$. After this step, we apply the bottleneck block with kernel size $1 \times 1 \times 1$ to the previous step output, and we get a feature map $M \in R^{c \times d \times n \times n}$. Here we omit the batch axis for simplicity. Let's denote $W = \{w_i\}_{i=1}^k$ to be a set of thicknesses for a maximum features aggregation, and $c$ — a central index of the second axis of feature map $M$. Thus, the output of the maximum feature aggregation block is $F = \big[\max\limits_{j \in [c-w_i;c+w_i]}M_{u,j,l,r}\big]_{i=1}^k$, $F \in R^{c \cdot k \times n \times n}$. Feature map $F$ can be used as input in a two-dimensional convolutional network. A self-attention block after the bottleneck layer can extend this design. Figure \ref{fig:mbmip} presents the scheme for the proposed blocks.

\subsection{Segmentation network with model-based feature projection blocks}
We propose a segmentation network architecture consisting of two blocks: a three-dimensional convolutional preliminary feature extractor and a U-Net-like primary segmentation model. A scheme of the proposed networks is presented in Figure \ref{fig:mbnet}. The feature extractor is a model-based 3D feature projection block on 2D plane. As generally segmentation networks can use any architecture without any restrictions, we decided to redesign UNet \cite{unet}. We changed the encoder to ResNest34 \cite{resnest} and ise dual dttention decoder blocks \cite{saunet}.  In addition to this base architecture, we propose a three-dimensional encoder model based on blocks of a 3D convolution layer with kernel size $3 \times 3 \times 3$, batch normalization, and ReLU activation. Each block $T_i$ consists of two 3D-blocks, with feature size $32 \cdot i$. $P_i$ is a model-based projection block with a bottleneck size of $4 \cdot i$. For correct subsampling, we use MaxPooling with strides $1 \times 2 \times 2$ after $T_i, i=1,2,3$. Outputs of blocks $P_i$ are concatenated with input features to blocks $E_i, i=1, 3, 4, 5$.

\section{Evaluation}
\label{sec:Evaluation}
\subsection{Experiment on LUNA2016}
\label{subsec:Experiment on LUNA2016}

\begin{figure}[tbp]
\centering
  \includegraphics[width=0.6\linewidth]{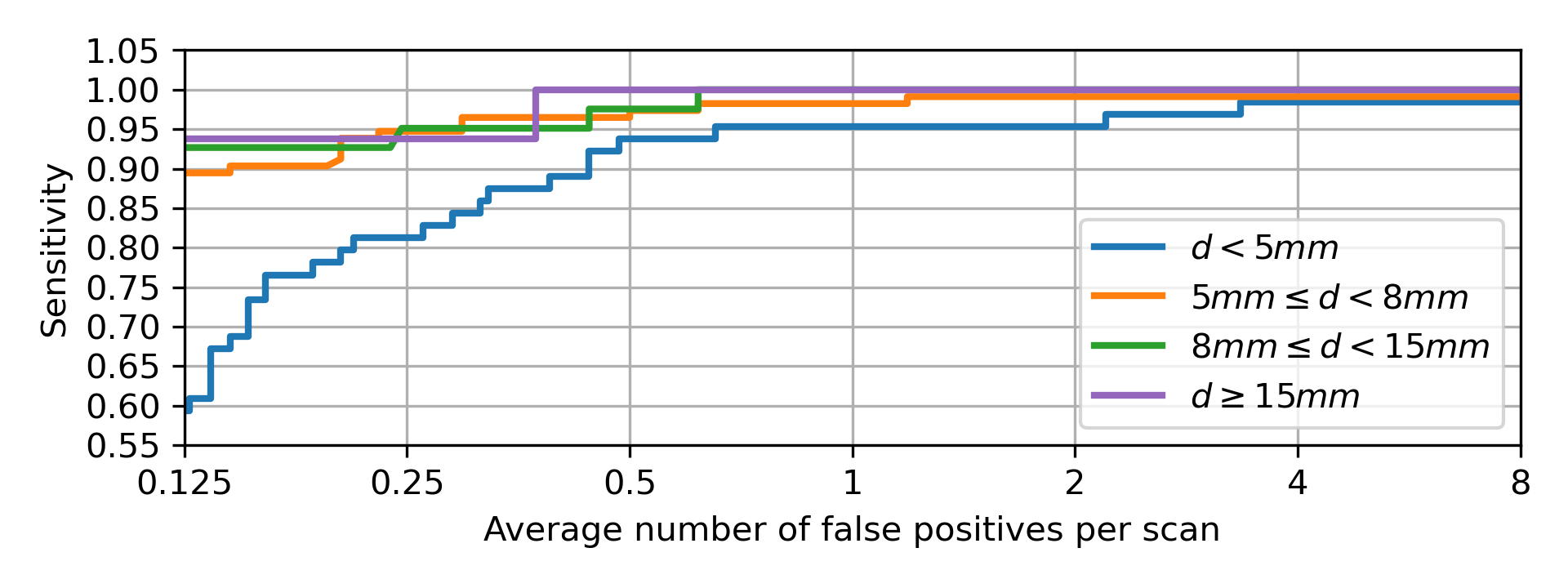}
  \caption{FROC curves depending on nodule size for AttentionMIP Enc.}
  \label{fig:diam_froc}
\end{figure}

For evaluation of the proposed framework, we use the commonly accepted baseline LUNA2016.
%We split all data with the following proportion: 65\% for the train, 10\% for the test, and 25\% for the evaluation. 
We resample CT scans to 0.8mm spacing and train networks with all three projections simultaneously. As input, we use a slab with a thickness of $16.8mm$ (21 slices). For $W = \{w_i\}_{i=1}^3$ we choose ${3, 6, 10}$ respectively. We choose the MinMax scaling function as a preprocessed function, perform a linear map of -1000Hu to 0 and 400Hu to 1, and clip all values outside this range. We use an Adam \cite{adam} optimizer with a learning rate of $0.0005$ and train each network for five epochs. We also use augmentation techniques: flips, rotations, and zoom of CT scans. We train four networks: 1) with maximum projection blocks, 2) with attention maximum projection blocks, 3) with a three-dimension parallel encoder with maximum projection blocks, and 4) with a three-dimension parallel encoder with attention maximum projection blocks. 

\begin{table}[htbp]
\centering
\addtolength{\tabcolsep}{-0.5pt}
  \caption{Results of the experiments on LUNA nodule detection track. The sensitivity per FP level per exam}
  \setlength{\tabcolsep}{2pt}
  \begin{tabular}{|c|c|c|c|c|c|c|c|c|}
  \hline
  \bfseries Experiment & \bfseries 0.125 FP & \bfseries 0.25 FP & \bfseries 0.5 FP & \bfseries 1 FP & \bfseries 2 FP & \bfseries 4 FP & \bfseries 8 FP & \bfseries Mean Sens\\\hline
  Li \cite{Experiments_7} & 0.739 & 0.803 & 0.858 & 0.888 & 0.907 & 0.916 & 0.920 & 0.862\\
  Wang \cite{Experiments_13} & 0.676 & 0.776 & 0.879 & 0.949 & 0.958 & 0.958 & 0.958 & 0.878\\
  Drokin \cite{Experiments_14} & 0.725 & 0.832 & 0.901 & 0.933 & 0.945 & 0.945 & 0.945 & 0.8894\\
  Khosravan \cite{Experiments_15} & 0.709 & 0.836 & 0.921 & 0.953 & 0.953 & 0.953 & 0.953 & 0.897\\
  Cao \cite{Experiments_17} & 0.848 & 0.899 & 0.925 & 0.936 & 0.949 & 0.957 & 0.960 & 0.925 \\\hline

  MaxMIP & 0.8 & 0.914 & 0.961 & 0.974 & 0.982 & 0.987 & 0.987 & 0.943\\
  AttentionMIP & 0.855 & 0.927 & 0.961 & 0.978 & 0.978 & 0.982 & 0.982 & 0.951\\
  MaxMIP Enc. & 0.855 & 0.936 & 0.961 & 0.978 & 0.978 & 0.982 & 0.982 & 0.953\\
  AttentionMIP Enc. & \textbf{0.872} & \textbf{0.931} & \textbf{0.965} & \textbf{0.978} & \textbf{0.987} & \textbf{0.991} & \textbf{0.991} & \textbf{0.959} \\\hline
  \end{tabular}
\label{tab:test2}
\end{table}

Table \ref{tab:test2} shows that the proposed framework with various segmentation networks outperforms recently published results.

In Figure \ref{fig:diam_froc}, we provide performance analysis for the best model depending on nodule size. The proposed model has a better performance at a nodule range with a diameter of more than 5 mm and has a tendency to display minor sensitivity at a diameter of less than 5 mm at low false-positives per scan rates.

\begin{figure}[htbp]
\centering
  \includegraphics[width=0.9\linewidth]{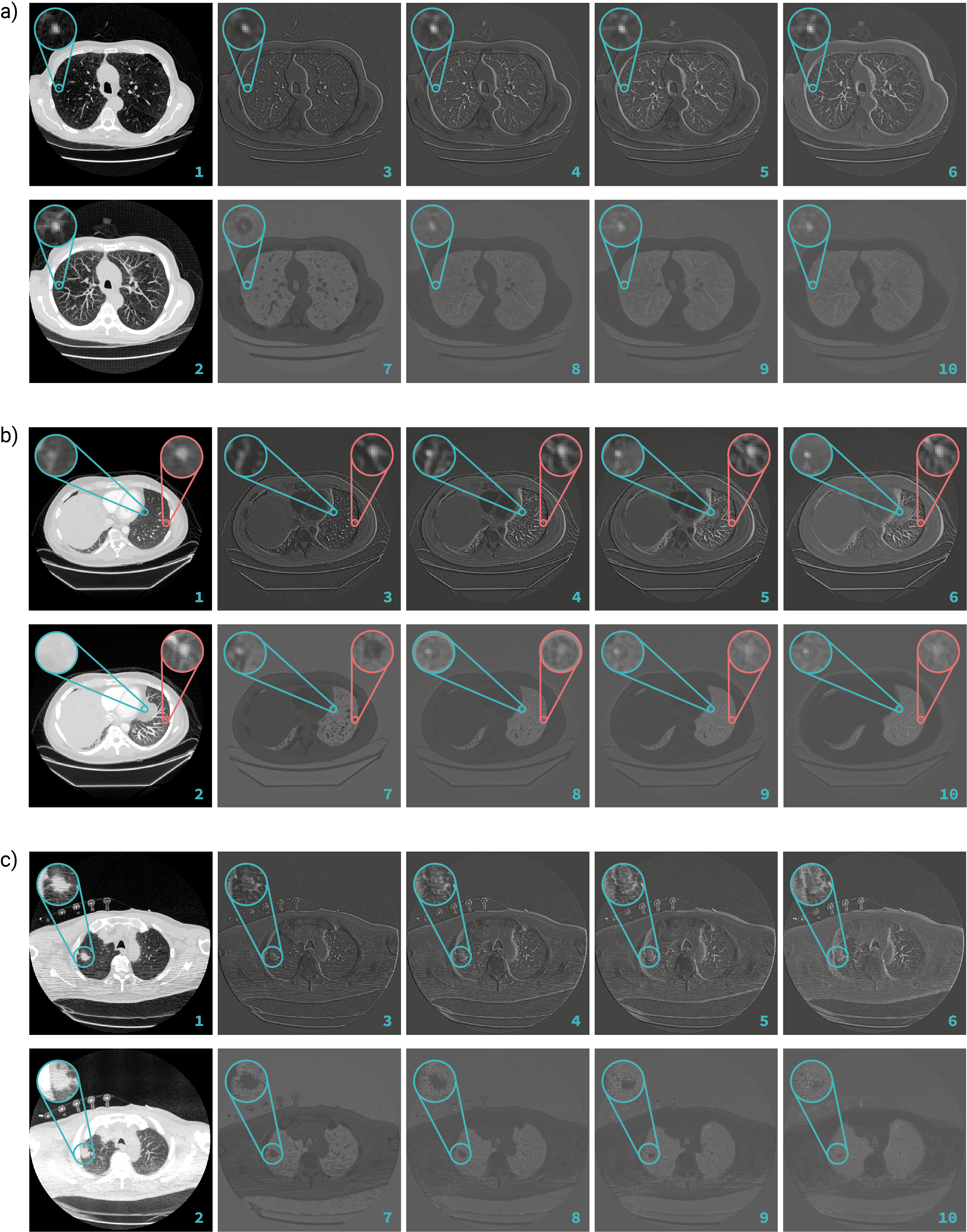}
  \caption{Feature visualization of proposed network. 1 - central slice; 2 is 21mm MIP on input data; 3 is a center slice of MaxMIP features, 4-6 is features of MaxMIP with $w_1=3, w_2=7, w_3=10$;7-10 same as 3-6, but for AttentionMIP. Circles denote true nodules.}
  \label{fig:one2rule}
\end{figure}

To understand the results better, we decided to add a visualization of the features fed to a two-dimensional segmentation network. Figure \ref{fig:one2rule} a. shows that model-based features tend to ignore normal lung tissue compared to MIPs images, to an extent. Nonetheless they preserve valuable information about lung nodules. Common false-positive detections include pieces of bronchi, blood vessels, fibrosis, etc. Due to this, the property of proposed model-based feature projection blocks allows models to reduce false-positives. Figure \ref{fig:one2rule} b. shows that the model projection block ignores diaphragmatic cupula, and Figure \ref{fig:one2rule} c. demonstrates that the proposed block is better at noise suppression of CT scanner than MIP projection.

\subsection{Ablation study}
\label{subsec:Ablation study}
As a part of the proposed method analysis, we include the results of an ablation study. We train and evaluate proposed models only on axial data. We also train a model and evaluate a framework with a "naive" block: we use the whole slab as a 21-channel input to proposed a 2D segmentation network without a model-based feature projection block. We train the model with a four-channel input to the segmentation model: the central slice and MIPs with a thickness equal to 5, 10, 15 mm, as was proposed in \cite{ml_mip}. Finally, we train the model with a maximum projection block and U-net \cite{unet} as a 2D segmentation network. Table \ref{tab:test4} shows the results of the ablation study. According to the provided experimental data, all proposed parts of the framework contribute positively to framework performance. As a consequence, the proposed innovations are not redundant.

\begin{table}[htbp]
\centering
\addtolength{\tabcolsep}{-0.5pt}
  \caption{Results of the ablation study on LUNA2016. The sensitivity per FP level per exam.}
  \setlength{\tabcolsep}{1pt}
  
  \begin{tabular}{|c|c|c|c|c|c|c|c|c|}
  \hline
  \bfseries Experiment & \bfseries 0.125 FP & \bfseries 0.25 FP & \bfseries 0.5 FP & \bfseries 1 FP & \bfseries 2 FP & \bfseries 4 FP & \bfseries 8 FP & \bfseries Mean Sens\\\hline
  \multicolumn{9}{|c|}{\bfseries Only axial data}\\\hline
  MaxMIP & 0.631 & 0.784 & 0.882 & 0.937 & 0.964 & 0.968 & 0.972 & 0.877\\
  AttentionMIP & 0.776 & 0.898 & 0.941 & 0.960 & 0.968 & 0.972 & 0.972 & 0.926\\
  MaxMIP Enc. & 0.768 & 0.858 & 0.964 & 0.968 & 0.972 & 0.972 & 0.972 & 0.925\\
  AttentionMIP Enc. & 0.784 & 0.901 & 0.933 & 0.949 & 0.968 & 0.976 & 0.984 & 0.928\\\hline
  \multicolumn{9}{|c|}{\bfseries Feature blocks}\\\hline
  MIP & 0.686 & 0.815 & 0.882 & 0.913 & 0.937 & 0.941 & 0.941 & 0.873\\
  Naive & 0.729 & 0.827 & 0.890 & 0.941 & 0.968 & 0.972 & 0.972 & 0.899\\\hline
  \multicolumn{9}{|c|}{\bfseries U-net}\\\hline
  MaxMIP & 0.749 & 0.862 & 0.942 & 0.972 & 0.980 & 0.987 & 0.987 & 0.925\\\hline
  \end{tabular}
\label{tab:test4}
\end{table}

\subsection{Real world data evaluation}
\label{subsec:realworldeval}

Our radiologist team analyzed 123 CT scans from Lomonosov Moscow State University Medical Research and Educational Center\footnote{https://www.msu.ru/en/info/struct/} and Yamalo-Nenets Autonomous Okrug, Russian Federation diagnosed with the proposed framework. These studies have a wide range of variability and include the analysis of verified malignancies, benign pathologies, solitary and multiple lesions, nodules of different size and localization, associations with other lung pathologies, series with various parameters (series with thin/thick slices, series with different reconstructions: lung or soft tissue, series with different protocols: standard Chest CT, low-dose Chest CT, Chest CT + contrast agent), absence/existence of breathing artifacts, etc. During the evaluation step, we followed the LIDC-IDRI markup protocol and we use include and exclude criteria described at LUNA2016 Challenge\footnote{https://luna16.grand-challenge.org/Data/}. This dataset contains 138 nodules. Proposed models achieve a sensitivity of $0.913$ at an FP level of $0.1869$ per scan.

\section{Conclusion and Discussion}
\label{sec:Conclusion and Discussion}
We have proposed a novel approach for lung nodule CAD systems based on the U-Net-like segmentation network with a model-based feature projection block as a preliminary feature extractor. Coupled with processing three CT scans projections, this novelty allows us to build a single-model approach and achieve SOTA results on LUNA2016. We have performed an extensive analysis of the proposed solution with the ablation study and the feature analysis. We also performed validation on the proposed model using external data. The results shows that the proposed method can be used for lung cancer screening and as a CADe system in real world setup. As the next step of our research, we plan to extend the proposed segmentation framework to other CT scan analysis cases. Moreover, maximum-based feature projection after a three-dimensional convolutional block can be suitable for video analysis, both in the medical and general domains.

\bibliographystyle{splncs04}
\bibliography{AIST_2020_bib}

\end{document}